\documentclass[pra,aps,final,twocolumn,nofootinbib,showpacs]{revtex4-1}
\usepackage{epsfig}
\usepackage{latexsym}
\usepackage{xspace}
\usepackage{hyperref}
\usepackage[latin2]{inputenc}
\usepackage{indentfirst}
\usepackage{enumerate}
\usepackage{color}
\usepackage{colordvi}

\usepackage{amsmath}
\usepackage{amssymb}
\usepackage[english]{babel}
\usepackage{url}

\newcommand{\eq}[1]{\begin{align} #1 \end{align}}

\def\be{\begin{equation}}
\def\ee{\end{equation}}
\def\bea{\begin{eqnarray}}
\def\eea{\end{eqnarray}}
\def\l{\label}

\def\d{\mbox{d}}

\def\siml{\;\hbox{\kern.1em \lower.7ex \hbox{$\sim$} \kern-1.12em
 \raise.5ex \hbox{$<$} \kern.1em}}
\def\simg{\;\hbox{\kern.1em \lower.7ex \hbox{$\sim$} \kern-1.12em
 \raise.5ex \hbox{$>$} \kern.1em}}

\def\siml{\;\hbox{\kern.1em \lower.7ex \hbox{$\sim$} \kern-1.12em
 \raise.5ex \hbox{$<$} \kern.1em}}
\def\simg{\;\hbox{\kern.1em \lower.7ex \hbox{$\sim$} \kern-1.12em
 \raise.5ex \hbox{$>$} \kern.1em}}

\begin{document}

\title{Effects of quantum statistics near the critical point of nuclear matter
}

\author{ S.N.\ Fedotkin and A.G.\ Magner}
\affiliation{Institute for Nuclear Research NASU, 03680 Kiev, Ukraine}

\author{M.I.\ Gorenstein}
\affiliation{
Bogolyubov Institute for Theoretical Physics, 03143 Kiev, Ukraine}

\begin{abstract}
  Effects of quantum statistics
  for nuclear matter equation of state are analyzed in terms of the recently
 proposed quantum van der Waals model.
 The system pressure is expanded
 over a small  parameter
  $\delta \propto n(mT)^{-3/2}[g(1-bn)]^{-1}$, where
$n$ and $T$ are, respectively, the particle number density and temperature,
$m$ and $g$ the particle mass and
 degeneracy factor.
 The parameter $b$ corresponds
 to the van der Waals excluded volume.
  The corrections due to   quantum statistics
  for the critical point values of
 $T_c$,
  $n_c$, and the critical pressure $P_c$
   are found within the linear and quadratic 
  orders over $\delta $.
 These approximate analytical results appear to be in a good agreement with exact numerical calculations
 in the quantum van der Waals model for interacting Fermi particles:
 the symmetric nuclear matter ($g=4$)
 and the pure neutron matter ($g=2$).
 They can be also applied to the system of interacting  Bose particles like  
the matter
composed of $\alpha$ nuclei.

\end{abstract}
%
%\pacs{24.10.Pa, 25.75.-q, 21.65.Mn}

\maketitle

\section{Introduction}
 A study of nuclear matter -- an interacting system of protons and neutrons --
has a long history; see, e.g.,
Refs.~\cite{nm-1,nm-2,nm-3,nm-4,nm-5,nm-6,nm-7,nm-8,nm-9,nm-10,nm-11}.
Realistic versions of the nuclear matter equation of state includes both
the attractive and repulsive forces
between protons and neutrons. A theoretical description of the
thermodynamical behavior of this matter
leads to the liquid-gas first-order phase transition
which ends at the critical point. Such a behavior is rather similar to
that in the atomic gases and liquids.
Experimentally, a presence of the liquid-gas phase transition in nuclear matter was  reported and then analyzed
in numerous of papers
(see, e.g., Refs.~\cite{ex-1,ex-2,ex-3,ex-4,ex-5,ex-6}).

In the present paper the recently proposed  van der Waals equation of state with  the quantum statistics~\cite{marik}
is used to describe the
properties of nuclear matter.
The aim of our consideration is to investigate the role and size of the effects of quantum statistics for the nuclear matter
properties near its critical point.
Our consideration will be restricted to small temperatures, 
$T \siml 30$~MeV, and not too large
particle densities.  Within these restrictions, the
number of nucleons
becomes a conserved number,
and the chemical potential of this system regulates the number density of nucleons.
We do not include the Coulomb forces
and make no differences between protons and neutrons (both these particles are named as nucleons).
In addition, under these restrictions the
non-relativistic treatment becomes very accurate
and is adopted in our studies.

The paper is organized as the following. In Sec.~\ref{sec-2} we remind some results of the
ideal Bose and Fermi gases. In Sec.~\ref{sec-3} the quantum  statistics effects
near the critical point are studied for the symmetric nuclear matter. These results are then extended to the pure neutron matter
and pure $\alpha$ matter. Section \ref{sec-5} summarizes the paper.

\section{Ideal quantum gases}\label{sec-2}
The pressure $P(T,\mu)$ plays the role of the thermodynamical potential in the grand canonical ensemble (GCE)
where temperature
$T$ and chemical potential $\mu$ are independent variables.
The particle number density
$n(T,\mu)$, entropy density $s(T,\mu)$, and energy density $\varepsilon(T,\mu)$  are given as
\eq{\label{term}
n=\left(\frac{\partial P}{\partial \mu}\right)_T~,~ s=\left(\frac{\partial P}{\partial T}\right)_\mu~,~
\varepsilon= Ts+\mu n-P~.
}
We start with the
GCE expressions for the pressure $P_{\rm id}(T,\mu)$
and particle number density $n_{\rm id}(T,\mu)$ for the ideal non-relativistic
quantum gas \cite{G,LLv5},
\eq{\l{Pid}
& P_{\rm id}=\frac{g}{3}\int \frac{\d {\bf p}}{(2\pi \hbar)^3}\frac{p^2}{m}
\left[\exp\left( \frac{p^2}{2mT} -\frac{\mu}{T }\right) - \theta\right]^{-1}~,\\
& n_{\rm id}=g\int \frac{\d {\bf p}}{(2\pi \hbar)^3}
\left[\exp \left( \frac{p^2}{2mT} -\frac{\mu}{T} \right) - \theta\right]^{-1}~,\label{nid}
}
where $m$ and $g$
are, respectively, the particle mass and degeneracy factor. The value of $\theta=-1$ corresponds to the Fermi gas,
$\theta=1$ to the Bose gas, and $\theta=0$ is the Boltzmann (classical) approximation when
effects of the quantum statistics are neglected \footnote{The units
  with Boltzmann  constant
    $\kappa_{\rm B}=1$ are used. We keep the Plank constant in the
  formulae  to illustrate
  the effects of quantum statistics, but put $\hbar=h/2\pi=1$ in all
  numerical calculations.  For simplicity, we omitted here and below the
 subscript $id$ for the ``ideal gas`` everywhere where
it will not lead to a misunderstanding. }.

Equations (\ref{Pid}) and (\ref{nid}) can be expressed in terms of the
power series over fugacity $z \equiv \exp(\mu/T)$ for $z\le 1$:
\eq{\label{Pid-1}
  P(T,z) &
   \equiv  \frac{gT}{\theta \Lambda^3}\,{\rm Li}_{5/2}(\theta z) =
  \frac{gT}{\theta \Lambda^3}\,\sum_{k=1}^\infty
  \frac{(\theta z)^k}{k^{5/2}}~ ,\\
  n(T,z) &
  \equiv
    \frac{g}{\theta \Lambda^3}\,{\rm Li}_{3/2}(\theta z)
     =  \frac{g}{\theta \Lambda^3}\,\sum_{k=1}^\infty \frac{(\theta z)^k}{k^{3/2}}~ ,
    \label{nid-1}
}
where
\be\l{lambdaT}
\Lambda~\equiv ~\hbar\sqrt\frac{2 \pi}{mT}~.
\ee
is the de Brogle heat wavelength  \cite{LLv5},
and $\mbox{Li}_\nu$
is the polylogarithmic function \cite{Grad-Ryzhik,Li} .
The values of $\mu >0$, i.e. $z>1$, are forbidden
in the ideal Bose  gas. The point
$\mu=0$ corresponds to an onset of the Bose-Einstein condensation in the system
of bosons. For fermions, any values of $\mu$
are possible, i.e., integrals (\ref{Pid}) and (\ref{nid}) exist for $\theta=-1$ at all real values of $\mu$.
However, the power series
(\ref{Pid-1}) and (\ref{nid-1}) are convergent at $z\le 1$ only. For the Fermi statistics at $z>1$, the integral
representation of the  corresponding polilogarithmic function
can be used.
Particularly,
at $z\rightarrow \infty$ one can use the asymptotic Sommerfeld expansion of the
$\mbox{Li}_{\nu}(-z)$ functions over $1/\mbox{ln}^2|z|$
\cite{brack}.

\begin{figure}
\begin{center}
  \includegraphics[width=7.0cm,clip]{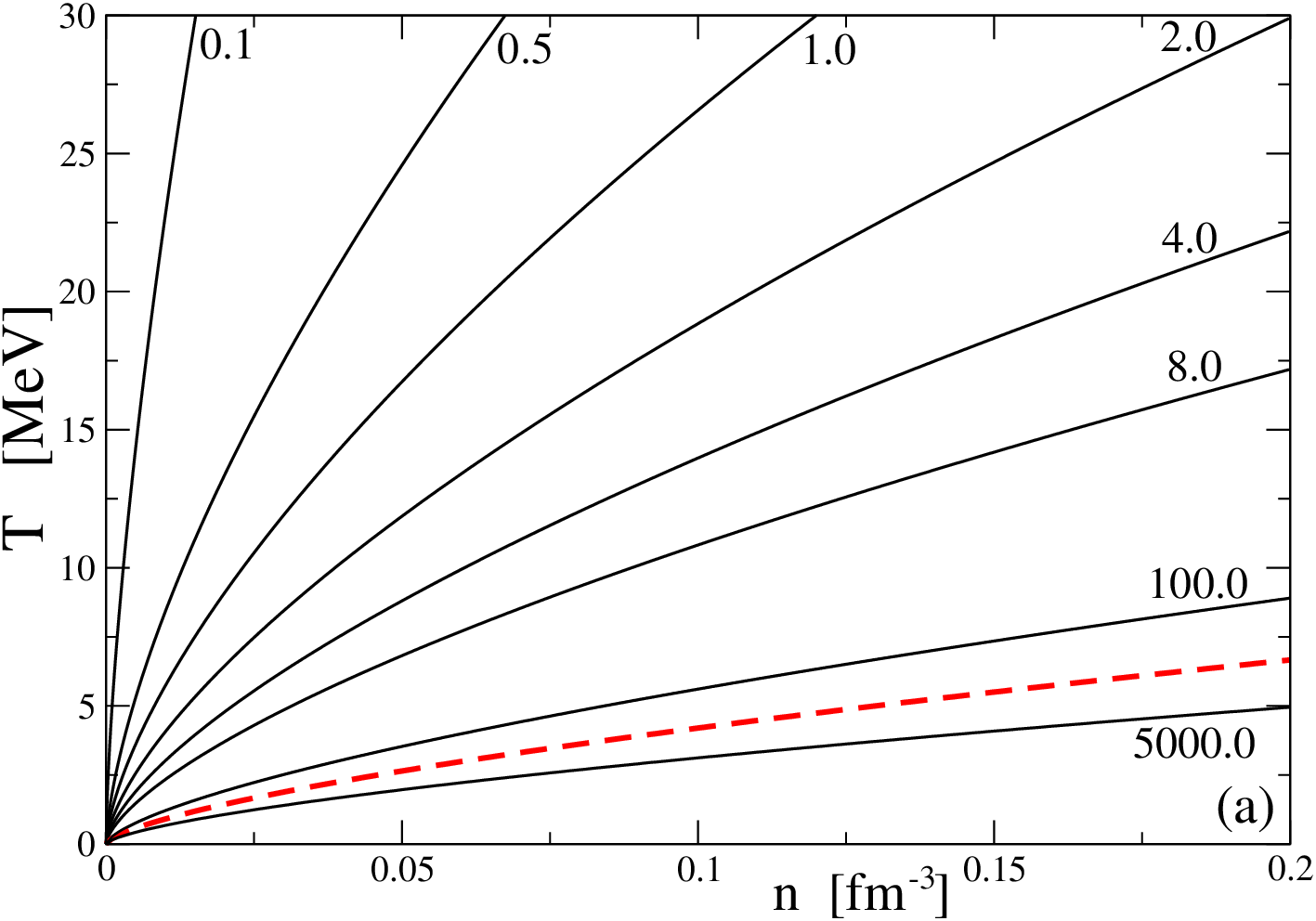}
 \includegraphics[width=7.0cm,clip]{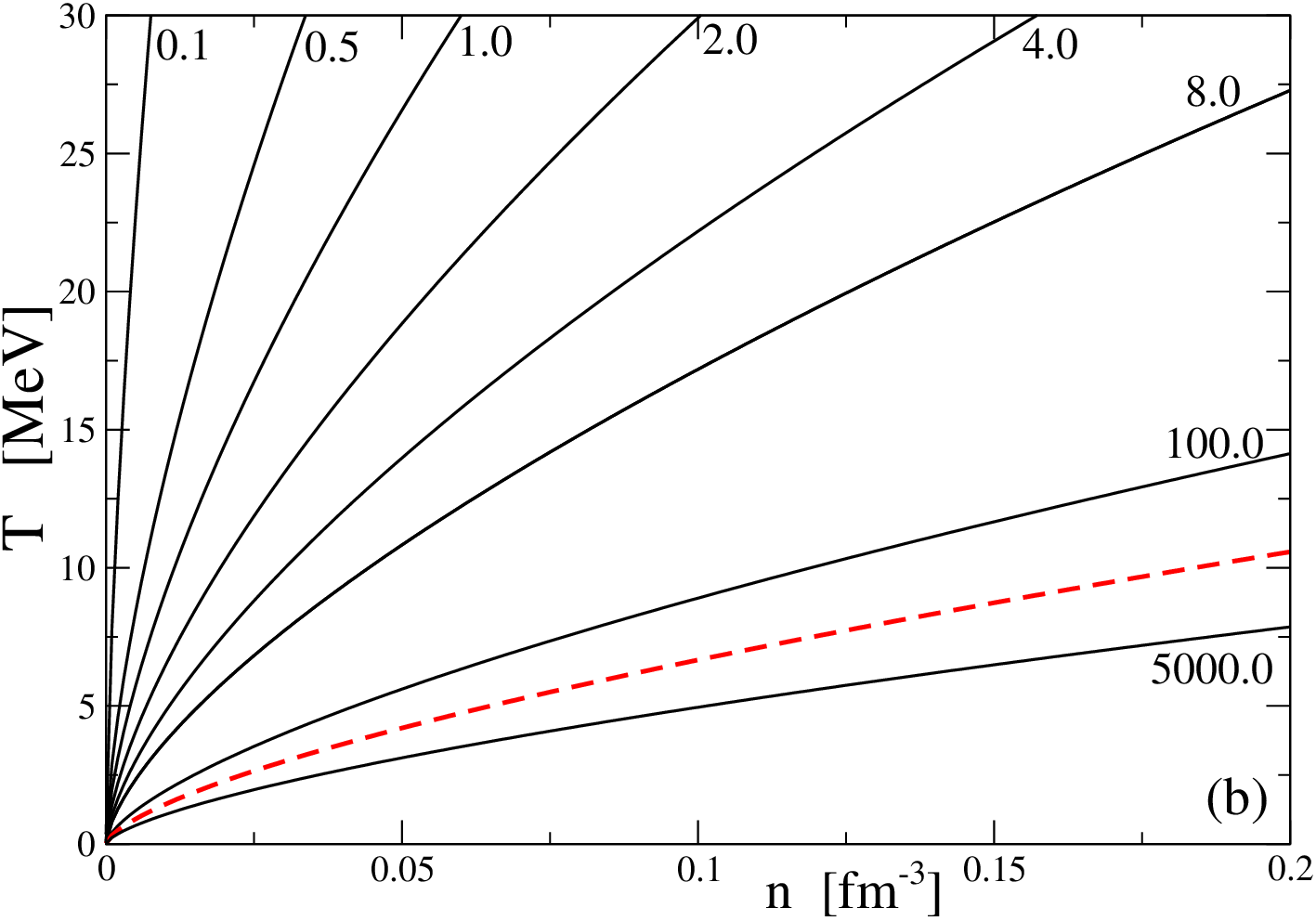}
 \includegraphics[width=7.0cm,clip]{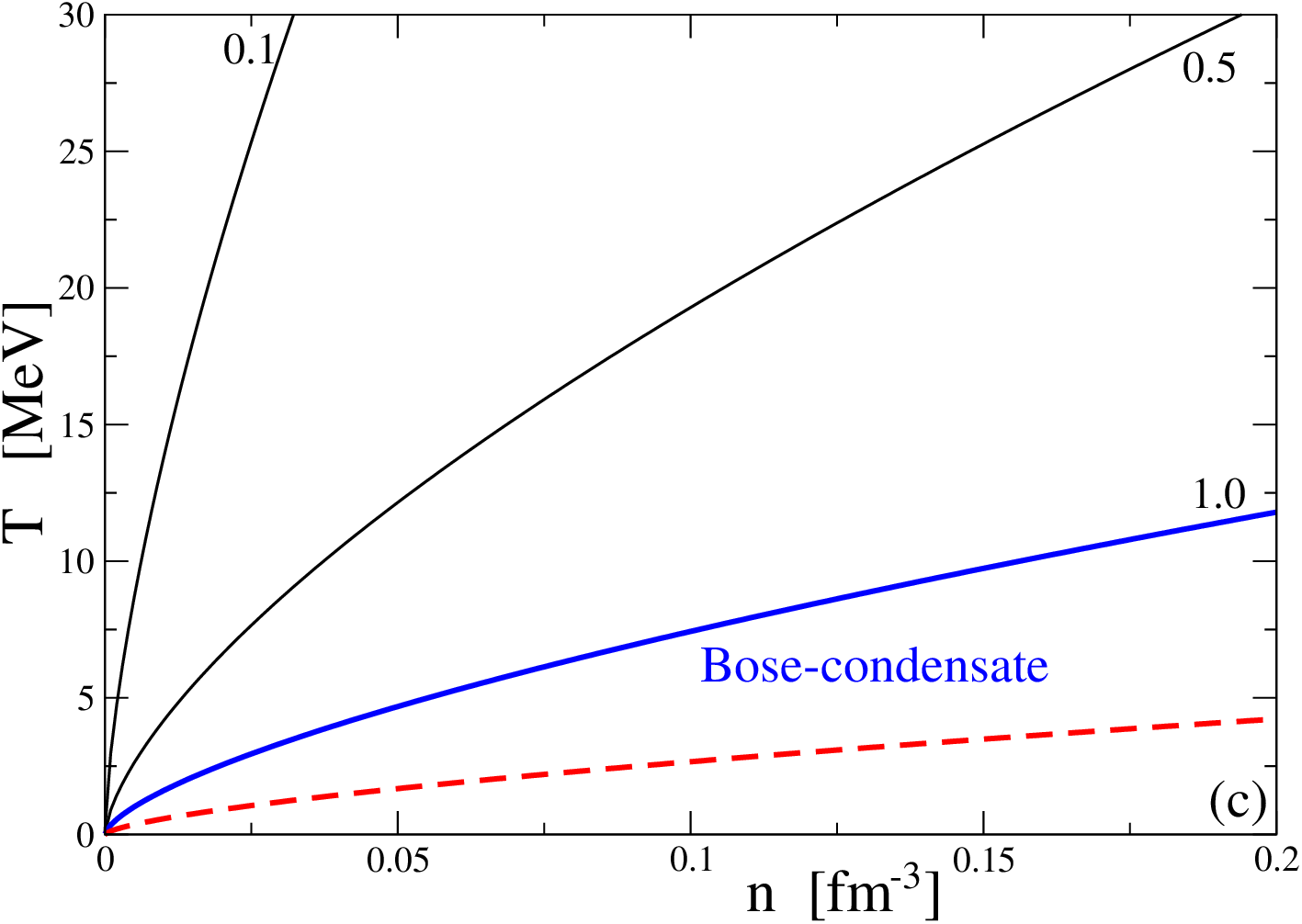}
\end{center}
\caption{
    Lines of the fixed fugacities $z=z(n,T)$
 are shown for the ideal Fermi gas of nucleons
  (a) and neutrons (b), and the ideal Bose gas of alpha particles (c).
  Dashed red lines correspond to $s=0$ for the entropy density
  (\ref{s}) of the classical ideal gas.
  A blue line in (c) shows the largest
  value of $z=1$ in the ideal Bose gas. This is the line
  of the Bose-Einstein condensation, and the  Bose condensate of
    $\alpha $ particles exists below this line.
 }
\label{fig-1}
\end{figure}

Figure \ref{fig-1} shows lines of the constant
values of fugacity $z$ in the $n$-$T$ plane
for the ideal Fermi gases (a) and (b), and
Bose  gas (c), see Eq.~(\ref{nid-1}).
Figures 1  (a)
corresponds to the isospin symmetric ideal nucleon gas
(i.e., an equal number
of protons and neutrons).
We take $m \cong 938$~MeV neglecting
a small difference between proton and neutron masses.
The degeneracy factor is then  $g=4$ which takes into account two spin and two isospin
states of nucleon. The ideal neutron gas with $g=2$ is presented in Fig. 1 (b),
and the ideal Bose gas of $\alpha$-nuclei ($g=1$ and $m\cong 3727$~MeV) is shown in Fig. 1 (c).

At $z\ll 1$,  only one
term $k=1$ is enough  in Eqs.~(\ref{Pid-1}) and (\ref{nid-1}) which leads
to the classical ideal gas relation
\eq{\label{Pid-cl}
P=n\,T~.
}
Note that the result (\ref{Pid-cl}) follows automatically from
Eqs.~(\ref{Pid}) and (\ref{nid})
at $\theta=0$.
As seen from Fig.~\ref{fig-1}, the classical Boltzmann approximation at $z\ll 1$
is valid for large $T$ and small $n$ region of the $n$-$T$ plane.
However, the classical ideal gas equation (\ref{Pid-cl}) becomes wrong
at small $T$ and/or large $n$.
The entropy density of the classical ideal gas reads
\be\label{s}
s= \left(\frac{\partial P}{\partial T}\right)_\mu =
n~\left(\frac{5}{2}~-~\frac{\mu}{T}\right)~,
\ee
and it becomes negative at $\mu/T> 5/2$. The particle number density
in the classical ideal gas at $\mu/T=5/2$ equals to
\eq{\label{n-cl}
n=\frac{g}{\hbar^3}\left(\frac{mT}{2\pi}\right)^{3/2}\exp \left(\frac{5}{2}\right)~.
}
The relation (\ref{n-cl})
is shown by a dashed red line
in Fig.~\ref{fig-1}. Under this line, the entropy density of the classical ideal gas
becomes negative.
This happens  at small $T$ and/or large $n$ and means
a contradiction with the third law of thermodynamics.
Quantum statistics solves this problem:
Eqs.~(\ref{Pid}) and (\ref{nid})
guarantee $s\ge 0$ at all $T$ and $n$, and $s=0$
at $T=0$.

%%%%% FIG.1 P(n,T).

Inverting the $z^k$ power series in  Eq.~(\ref{nid-1}), one finds
the power expansion of $z$ over the parameter $\epsilon$
(see, e.g., Ref.~\cite{balescu}),
\eq{\label{eps}
  \epsilon\equiv -\frac{\theta n \Lambda^{3}}{4\sqrt2\, g}\equiv
  -\theta\, \frac{ \hbar^3\,\pi^{3/2}~n}{2\,g\,(mT)^{3/2}}~.
}
This expansion is inserted then to Eq.~(\ref{Pid-1}).
At small $|\epsilon | <1$ the expansion of the pressure over
the powers of $\epsilon$
is rapidly
convergent,
and a few first terms give already a good approximation of the
quantum statistics effects.
Taking   the three
terms, $k=1$, $2$, and $3$,
in
Eqs.~(\ref{Pid-1}) and (\ref{nid-1}),
one obtains a classical gas result (\ref{Pid-cl})
plus the  corrections due to the effects of quantum statistics:
\eq{\label{Pid-n}
P(T,n)= n T \left[1 +
  \epsilon - c_2^{}\epsilon^2 +
  \mbox{O}(\epsilon^3)\right]~,
}
where
$c^{}_2=4[16/(9\sqrt{3})-1] \cong 0.106$~.
For brevity, we call the linear and quadratic $\epsilon$-terms in
  Eq.~(\ref{Pid-n})
as the first and second (order) quantum corrections.

Equation (\ref{Pid-n}) demonstrates explicitly a deviation of  the quantum ideal gas pressure
from the classical ideal gas  value (\ref{Pid-cl}): the Fermi statistics leads to an increasing of the classical pressure,
while the Bose statistics to its decreasing. This is often interpreted \cite{LLv5} as the effective
Fermi `repulsion' and Bose `attraction' between quantum particles.

\section{Quantum van der Waals model}\label{sec-3}

Recently, the van der Waals (vdW) equation of state
was  extended by taking into account the effects of quantum statistics
in Ref.~\cite{marik}.
The pressure function of the quantum vdW (QvdW) model can be presented as \cite{marik}
\eq{\label{PQvdW}
& P(T,n)=P_{\rm id}\left[T, n_{\rm id}(T,\mu^*)\right]
- an^2~, \\
& n_{\rm id}(T,\mu^*)~=~ \frac{n}{1-bn}~,\label{nvdw}
}
  where $P_{\rm id}$ and $n_{\rm id}$ are respectively given by  Eqs.~(\ref{Pid}) and
(\ref{nid}).
The  constants
$a>0$ and $b>0$ are responsible for respectively attractive and repulsive
interactions between particles.
The QvdW model
  given by Eqs.~(\ref{PQvdW}) and (\ref{nvdw})
 satisfies
the following conditions:

1. In the Boltzmann approximation, i.e. at $\theta=0$ in Eqs.~(\ref{Pid}) and (\ref{nid}), the QvdW
model (\ref{PQvdW}) and (\ref{nvdw}) is reduced to the {\it classical} vdW model
\cite{LLv5}
\eq{\label{vdW}
P=\frac{nT}{1-nb} - an^2~.
}
Note that the classical vdW model (\ref{vdW}) is further reduced
to the ideal
classical gas (\ref{Pid-cl}) at $a=0$ and $b=0$.

2. At $a=0$ and $b=0$ the QvdW model (\ref{PQvdW}) and (\ref{nvdw})
is transformed to the  {\it quantum} ideal gas  Eqs.~(\ref{Pid}) and (\ref{nid}).

3. The QvdW model (\ref{PQvdW}) and (\ref{nvdw}), in contrast to its classical version (\ref{vdW}), satisfies
the third law of thermodynamics by having a
  non-negative entropy with $s= 0$  at $T=0$.

We fix the model parameters $a$ and $b$ using the ground state
properties  of  the symmetric nuclear matter
(see, e.g., Ref.~\cite{nm}): at $T=0$ and
  $n=n_{0}= 0.16~\mbox{fm}^{-3}$, one requires
$P=0$ and the binding energy per nucleon $\varepsilon(T=0,n=n_0)/n_0=- 16$~MeV.
With the step-like Fermi momentum distribution for nucleons
the  analytical expressions for the thermodynamical quantities at $T=0$ are
obtained. One then finds from the above requirements\footnote{The multi-component
QvdW model with different $a$ and $b$ parameters for protons and neutrons was
discussed in Refs.~\cite{vova,roma}}:
\be\l{ab}
a=329.8\, \mbox{MeV} \cdot \mbox{fm}^3 , \;\;\; b=3.35\, \mbox{fm}^3~.
\ee
These values are very close to those
found in  Ref.~\cite{marik}.
Small differences appear  because of  the
non-relativistic formulation used in the
present studies.

In what follows, the first and second quantum correction
of the QvdW model
will be considered.
This
is analogous to that in Eq.~(\ref{Pid-n}) for the ideal quantum gas.
  Expanding $P_{\rm id}\left[T, n_{\rm id}(T,\mu^*)\right]$
in  Eq.~(\ref{PQvdW})
over the  small parameter $\delta$ (with $\theta=-1$ for
fermions and $\theta=1$ for bosons),
\eq{\label{del}
\delta = \frac{\epsilon}{1-bn}~=~
  -\theta\, \frac{ \hbar^3\,\pi^{3/2}~n}{2\,g\,(1-bn)(mT)^{3/2}}~,
}
one obtains
\eq{\label{Pvdw-n}
 P(T,n) = \frac{nT}{1-bn}\,\left[1~+~\delta
  ~-~~c^{}_2 \delta^2~+~
    \mbox{O}\left(\delta ^3\right) \right]
   -~ a\,n^2~,
}
   where $c^{}_2$ is the same small number coefficient
     as in Eq.~(\ref{Pid-n}).
     Similarly to the ideal gases, the quantum correction in
     Eq.~(\ref{Pvdw-n}) increases
with the particle number density $n$ and decreases with the system temperature $T$,
particle mass $m$, and degeneracy factor $g$. A new feature of the quantum
effects in the system of particle with the vdW interactions is the additional
factor $(1-bn)^{-1}$ in the quantum correction $\delta$, i.e., the
quantum statistics
effects becomes stronger due to the repulsive interactions between particles.

The vdW model, both in its classical form (\ref{vdW}) and in its QvdW
extension (\ref{PQvdW}) and (\ref{nvdw}),
describes the first order liquid-gas
phase transition. The critical point (CP) of this transition satisfies the following
equations:
\eq{\label{CP-0}
\left(\frac{\partial P}{\partial n}\right)_T = 0~,~~~~
\left(\frac{\partial^2 P}{\partial n^2}\right)_T=0~.
}
Using Eq.~(\ref{Pvdw-n}) in the first approximation in $\delta$,
one derives
from Eq.~(\ref{CP-0})
the
system of two equations
for the CP
parameters $n_c$ and $T_c$ 
at the same first order:
\bea\l{cp-1}
&&\frac{T}{2an\left(1-nb\right)^2}\left[1+ 2\delta\right]~=~1~,\\
&&\frac{bT}{a\left(1-nb\right)^3}
\left[1+ \frac{(1+2nb)}{b\,n}\,\delta\right]~=~1~. \label{cp-2}
\eea
Solving the system (\ref{cp-1}) and  (\ref{cp-2}),
one finds
in the same first order approximation
over $\delta$:
\eq{
 & T_c^{(1)} ~ \cong
  T_c^{(0)}\left(1 - 2 \delta_0  \right)
 \cong 19.0~{\rm MeV}~,\nonumber \\
& n_c^{(1)} \cong  n^{(0)}_c\left(1 - 2\delta_0  \right)
\cong 0.065~{\rm fm}^{-3}~.  \label{nc-1}
}
In Eq.~(\ref{nc-1}),
\eq{
  & T_c^{(0)}=\frac{8a}{27b}\cong 29.2~{\rm MeV}~, ~~~
  n_c^{(0)}=\frac{1}{3b}\cong 0.100~{\rm fm}^{-3}~,\nonumber \\
& P_c^{(0)}=\frac{a}{27b^2}\cong 1.09~{\rm MeV}\cdot {\rm fm}^{-3}~ \label{CP}
}
are the CP parameters of the classical vdW model, i.e., they are found from Eq.~(\ref{CP-0})
for the equation (\ref{vdW}).
The parameter $\delta_0$ in Eq.~(\ref{nc-1}) is given by Eq.~(\ref{del})
calculated at the CP (\ref{CP}), i.e. at $n=n_c^{(0)}$ and $T=T_c^{(0)}$.
Substituting Eq.~(\ref{nc-1}) for the critical temperature
 $T_c^{(1)}$ and density $n_c^{(1)}$
into Eq.~(\ref{Pvdw-n}) at the same first order,
for the CP pressure $P_c$
one obtains
\eq{\label{Pc-1}
  P_c^{(1)}
 \cong 0.48~{\rm MeV}\cdot {\rm fm}^{-3}~.
}

The numerical calculations within the full
QvdW model (\ref{PQvdW})
and (\ref{nvdw}) give
\eq{
& T_c\cong 19.7~{\rm MeV}~, ~~~n_c\cong 0.072~{\rm fm}^{-3}~,\nonumber \\
& P_c\cong 0.52~{\rm MeV}\cdot {\rm fm}^{-3}~. \label{CP-num}
}
These our results
(\ref{CP-num})
appear to be essentially the same as
those  obtained in Ref.~\cite{marik}.

 A summary of the results for the CP parameters
is presented in Table \ref{table-1}.
A difference of the results for the  classical vdW model (\ref{CP}) and QvdW model
(\ref{CP-num})
demonstrates
a role of the effects of Fermi statistics at the CP of the symmetric nuclear matter. The size of
these effects appears to
be rather significant. On the other hand, it is remarkable that the  first order correction
(\ref{nc-1}) and (\ref{Pc-1}) reproduce these effects of quantum statistics with  a high accuracy.  The second
order correction in the
expansion (\ref{Pvdw-n})
leads to an improvement of the
CP  parameters, $T_c^{(2)}=19.7$~ MeV, $n_c^{(2)}=0.072$~ fm$^{-3}$ and
$P_c^{(2)}=0.53$~
MeV$\cdot$fm$^{-3}$, which become closer to the numerical results
(\ref{CP-num}) of the full QvdW model.

%%\ee
%%%%%%%%%%%%% TABLE 1 %%%%%%%%%%%%%%%%%%%%%%%%%%%%
\vspace{0.3cm}
\begin{table}[pt]
\begin{center}
\begin{tabular}{|c|c|c|c|}
\hline
Critical points  &   Eq.(\ref{CP})  & First order correction & QvdW  \\
&       & Eqs.(\ref{nc-1}) and (\ref{Pc-1}) & Eq.(\ref{CP-num})  \\
\hline
$T_c$~[MeV] & ~29.2~& ~19.0~
& ~19.7~~\\
\hline
$n_c$~[fm$^{-3}$] &0.100 & 0.065
& 0.072
\\
\hline
$P_c$~[MeV$\cdot$ fm$^{-3}$] & 1.09  & 0.48
& 0.52
\\
\hline
\end{tabular}
\vspace{0.2cm}
\caption{{\small
Results for the CP parameters of the symmetric nuclear
matter ($g=4$).
}}
\label{table-1}
\end{center}
\end{table}
%%%%%%%%%%%%%%%%%%%%%%%%%%%%%%%%%%%%%%%%%

\vspace{0.3cm}
\begin{table}[pt]
\begin{center}
\begin{tabular}{|c|c|c|c|}
\hline
Critical points &
 Eq.(\ref{CP})  & First order correction  & QvdW  \\
\hline
$T_c$~[MeV] & ~29.2~&~ 8.7~~
& ~10.8~
\\
\hline
$n_c$~[fm$^{-3}$] &~0.100~&~0.030~
&~0.051
\\
\hline
$P_c$~[MeV$\cdot$ fm$^{-3}$] & ~1.09~  & ~0.13~
&~0.20~
\\
\hline
\end{tabular}
\vspace{0.2cm}
\caption{{\small
Results for the CP parameters for the neutron
matter ($g=2$).
}}
\label{table-2}
\end{center}
\end{table}

The CP parameters of the classical vdW model (\ref{CP}) depend  on the interaction
parameters $a$ and $b$, and they are not sensitive to the values of particle mass $m$ and degeneracy factor $g$.
The effects of quantum statistics change this conclusion.
The parameter  $\delta$
given by Eq.~(\ref{del}) is proportional to $m^{-3/2}\cdot g^{-1}$,
i.e., the effects of quantum statistics  become smaller for heavier
particles (e.g., for the light nuclei admixture in the nuclear matter)
or for large values of degeneracy factor $g$.
Particularly,
for $g\gg 1$ the quantum statistics  effects become negligible,
and the QvdW model (\ref{PQvdW},\ref{nvdw})
is reduced at $T>0$ to the classical vdW model (\ref{vdW}).
These results are in agreement with the numerical calculations in
the recent paper \cite{roma}.

The effects of quantum statistics become larger in the neutron matter
within the QvdW model with the same $a$ and $b$ parameters, but $g=2$ instead of $g=4$ for the symmetric nuclear matter.
This leads to essentially larger effects of the quantum statistics and stronger changes of the CP parameters.
These results are summarized\footnote{The properties of the pure neutron matter appear 
to be very sensitive to the values of the vdW parameters $a$ and $b$ for
neutrons. This is discussed in Ref.~\cite{roma}}  in Table \ref{table-2}.

\begin{figure}
\begin{center}
   \includegraphics[width=7.0cm,clip]{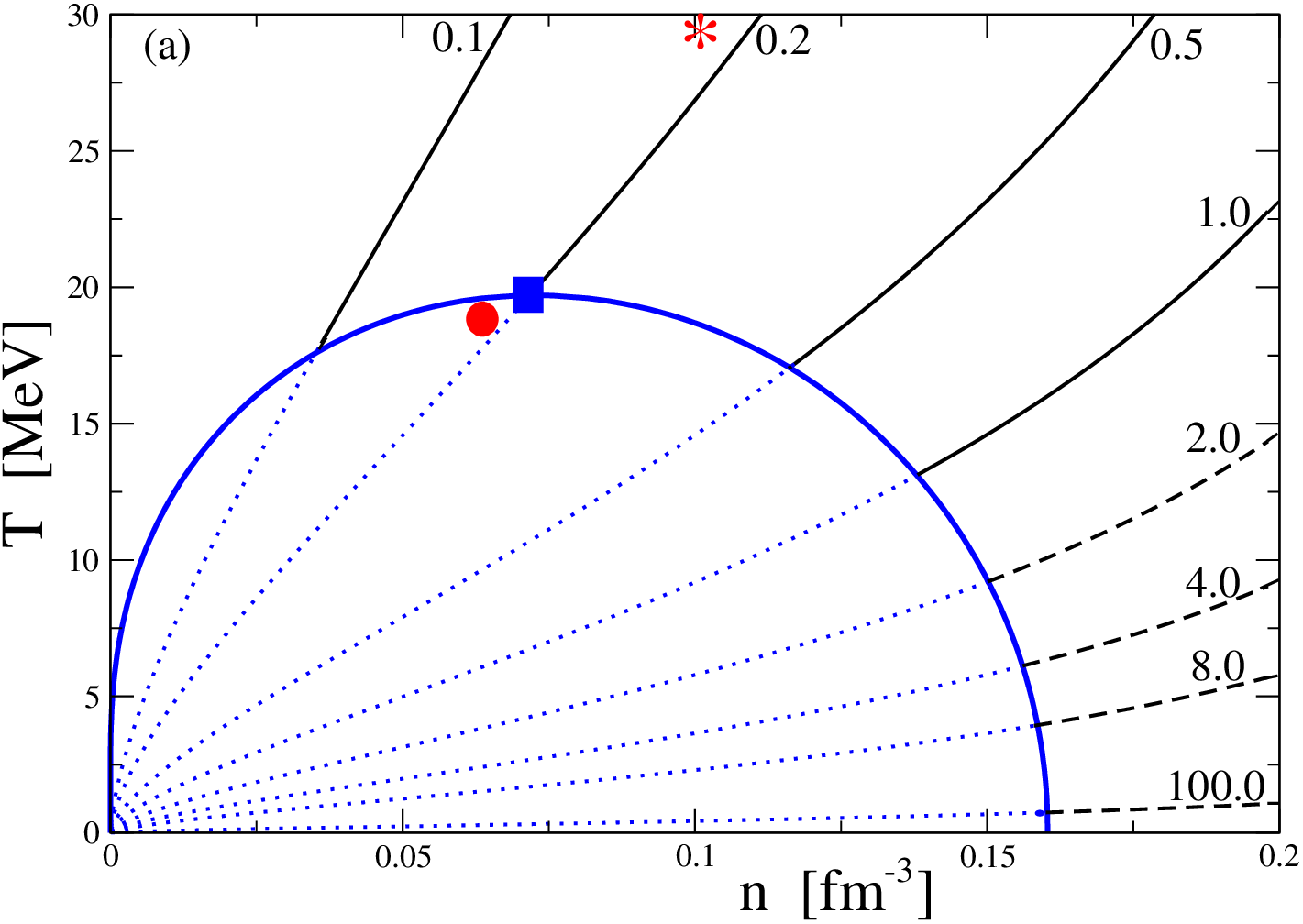}
   \includegraphics[width=7.0cm,clip]{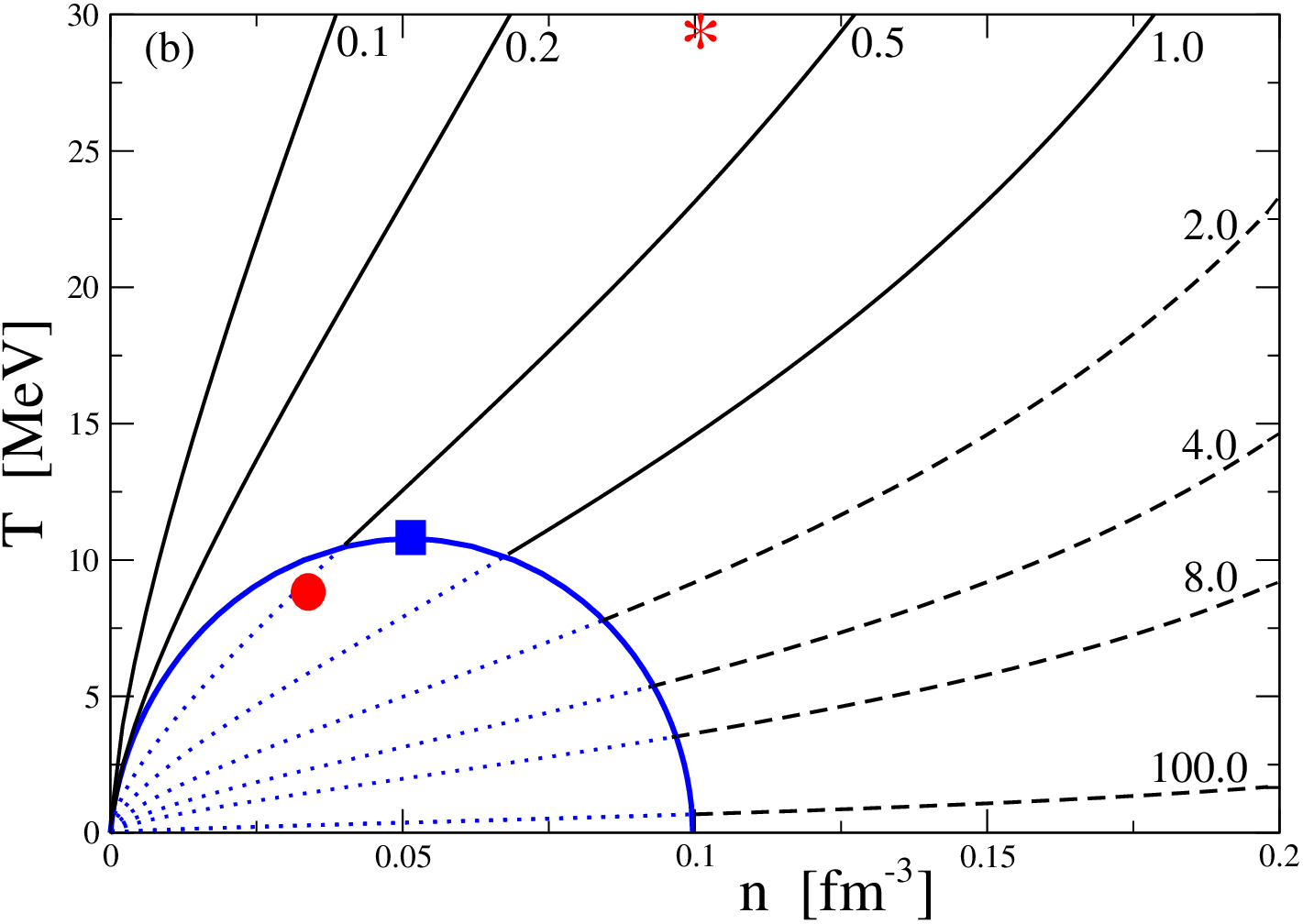}
    \includegraphics[width=7.0cm,clip]{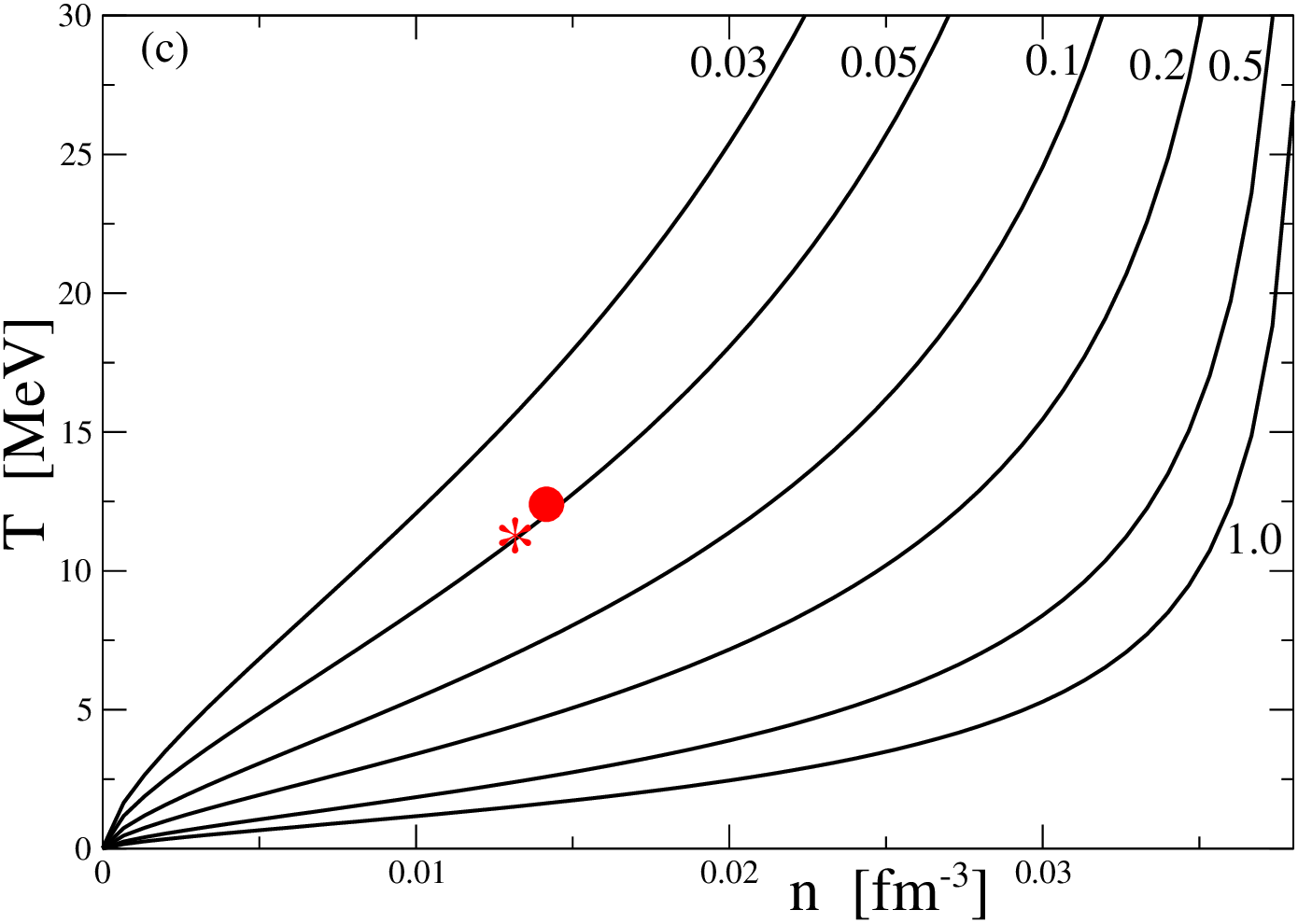}
\end{center}
\caption{
  Lines of the constant values of the parameter $\delta$
 [Eq.~(\ref{del})] in the $T$-$n$ plane
    for the symmetric nuclear matter  (a),  neutron matter (b),
    and $\vert \delta \vert$ for  $\alpha$-matter (c).
 The boundaries of the liquid-gas mixed phase regions
    are shown in (a) and  (b) by solid blue lines.
    The red close dots, blue squares, and asterisks show the  CP
    parameters in the respectively QvdW model within the first order correction,
    the full QvdW model, and classical vdW model.
  }.
\label{fig-2}
\end{figure}
In Fig.~\ref{fig-2} the expansion parameter $\delta$
entered to Eq.~(\ref{Pvdw-n})
is shown in the $n$-$T$ plane in the QvdW model for the symmetric nuclear matter
  (a)  and pure  neutron matter (b),   and the parameter $~-\delta~$  for the pure  $\alpha$  matter (c).
For the symmetric nuclear matter in Fig.~\ref{fig-2} (a) the values
of $\delta$  in a vicinity of the CP
is rather
 small,
$\delta \approx 0.2$. This explains a good accuracy of the first order
corrections of the quantum statistics effects.

For the neutron matter shown in Fig.~\ref{fig-2} (b) the
parameter of the quantum correction is $\delta\approx 0.5$ near the CP.
The correction due to the quantum statistics becomes indeed larger than that for the
symmetric nuclear matter.
This also explains why an agreement of the
first-order quantum results for the CP parameters with their exact
QvdW numerical values is worse than in the case of symmetric nuclear matter.
The second order correction in Eq.~(\ref{Pvdw-n}) leads to the critical values
$T_c^{(2)}=10.6$ MeV, $n_c^{(2)}=0.052$~fm$^{-3}$,
and $P_c^{(2)}=0.20$~MeV$\cdot$fm$^{-3}$. This improves
significantly an
agreement with results of the full QvdW model calculations presented in Table \ref{table-2}.

The calculations for interacting $\alpha$ particles within the QvdW model are shown in
Fig.~\ref{fig-2} (c).
For illustrative purposes we take the vdW parameters $a$ and $b$ of the pure $\alpha$-matter
to reproduce
the estimates
$T_c=11.2$~MeV 
and $n_c=0.013$~fm$^{-3}$ 
from Ref.~\cite{satarov} within  the classical vdW model.
One finds a rather small value
  $\vert \delta^{}\vert \approx 0.05$ for the $\alpha$ matter in a
vicinity of the CP, $\delta < 0$
[see Fig.~\ref{fig-2} (c)].
Note that the Bose and Fermi statistics  lead to the same
absolute values of the first order quantum correction (\ref{del}),
but with the different signs.
Thus, the quantum Bose effects  change the CP parameters into the opposite
direction,
i.e., $T_c$ and $n_c$ are moved
to larger
values due to the Bose effects  in comparison
with their classical vdW  values.
As seen from Fig.~\ref{fig-2} (c), there is  about a 10\%  increase  of the
$T_c^{(0)}$ and $n_c^{(0)}$
values due to the Bose statistics correction.

At small $T$ and/or large $n$,
the parameter $\delta \propto n (1-bn)^{-1}\cdot T^{-3/2}$
becomes large. In
this region of the phase
diagram, the QvdW model
should be treated within the full quantum statistical
formulation (\ref{PQvdW}) and (\ref{nvdw}).
In the $n$-$T$ region with $\delta > 1$ shown in
Fig.~\ref{fig-2} (a) and (b) by the
dashed lines  the first order quantum
approximation looses their meaning.

The mixed gas-liquid phases of the symmetric nuclear matter and pure
neutron matter
correspond to the $n$-$T$ regions under the blue solid lines
presented in Fig.~ \ref{fig-2} (a) and (b), respectively. The physical states
inside the mixed phase correspond to the equilibrium of the gas and liquid components
with equal values of $T$, $\mu$, and $P$. However, these components have different
particle number densities, $n_{\rm gas}< n_{\rm liq}$. They become very different,
$n_{\rm gas} \ll n_{\rm liq}$,
at small temperatures. In this case, the effects of quantum statistics
are small for the gas component ($\delta \ll 1$)  but are rather
large
($\delta \simg 1$)
  for the liquid one. For the Bose $\alpha$ particle system large
  quantum effects mean a possibility
of the Bose-Einstein condensation.
Thus, both effects -- the first-order phase transition
and a formation of the Bose condensate -- should be
treated simultaneously (see, e.g., Ref.~\cite{satarov}).
These questions are however outside of the scope of the present paper.

Rough
estimates give $\vert \delta^{}_0 \vert\ll 1$ in most cases of the CP
for different atomic gases \cite{kobeLynn53} and, thus,
small effects of quantum statistics.
This happens, despite of much smaller  values 
of the atomic critical
temperature $T_A\ll T_c$ in comparison with
the nuclear matter $T_c$ values.
The small effects of quantum statistics in the atomic systems in their
gas and liquid states
is a consequence  of {\it very small} atomic densities 
as compared to the nuclear ones,
$n_A \ll n_{c}$.  One exception from these arguments is the atomic system of He-4. The experimental value
for its CP parameters \cite{kobeLynn53,He},
$T_c({\rm He})\cong 5.2~$K$^0$ and $n_c({\rm He})\cong 10^{22}$~cm$^{-3}$,
lead to the estimate
$\delta^{}_0({\rm He})\approx -0.1$.
Therefore, our consideration within the first order quantum correction
over $\delta$ looks rather suitable to the analysis of the
quantum statistics effects
near the CP  of this atomic system.

\section{Summary}\label{sec-5}

The QvdW equation of state has been used to study
the quantum statistics effects
in a vicinity of the critical point of nuclear matter. To obtain the analytical
expressions,
the first order quantum statistics correction over the  small parameter $\delta$ is considered.
An explicit dependence on the system parameters is demonstrated.  Particularly, the CP position depends
on the particle mass $m$ and degeneracy factor $g$. Such a dependence is absent within the classical vdW model.
The quantum corrections to the CP parameters of the symmetric nuclear matter
appear to be quite significant.
For example, the value of $T_c^{(0)}=29.2$~MeV in the classical vdW model
{\it decreases} to the value $T_c^{(1)}=19.0$~MeV. On the other hand,
this approximate analytical  result within the first order
quantum correction is already  close to the numerical value of $T_c=19.7$~MeV
obtained by the numerical calculations within the full QvdW model.
The quantum correction of the CP parameters becomes even larger for the pure
neutron matter.
In this case, the classical value of $T^{(0)}_c=29.2$~MeV {\it decreases}
to $T_c^{(1)}=8.7$~MeV which
is still close to the numerical value
of the full QvdW model value $T_c=10.8$~MeV, 
  and the second order corrections
improves this agreement.

Our consideration is straightly  extended to the system of interacting bosons.
The quantum corrections
have   different signs
for fermions and bosons.
An example of the pure $\alpha$ matter has been considered.
The CP temperature $T_c^{(0)}=11.2$~MeV for the
the classical vdW model of  $\alpha$ particles
{\it increases} by about of 10\%  to
$T_c^{(1)}=12.3$~MeV for $\alpha$-matter  within the first order approximation
in  the QvdW model.

%%%%%%%%%%%%%%%%%%%%%%%%%%%%%%%%%%%%%%%%%%%%%%%%%%%%%%%%%%%%

%
\begin{acknowledgments}
We thank
D.V. Anchishkin,
B.I. Lev, A. Motornenko, R.V Poberezhnyuk, A.I. Sanzhur, and V. Vovchenko, for
many fruitful discussions.
The work of S.N.F. and A.G.M. on the project
``Nuclear matter in extreme conditions'' was
supported by the Program ``Fundamental researches in high energy physics
and nuclear physics (international collaboration)''
at the Department of Nuclear Physics and Energy of the National
Academy of Sciences of Ukraine.
The work of M.I.G. was supported
by the Program of Fundamental Research of the Department of
Physics and Astronomy of the National Academy of Sciences of Ukraine.
\end{acknowledgments}

\end{document}